\documentclass[preprint,longbibliography]{revtex4-1}

\usepackage[english]{babel}
\usepackage[utf8]{inputenc}
\usepackage{amsmath, amssymb,graphicx}
\usepackage[cdot,mediumqspace,squaren]{SIunits}
\usepackage{braket}
\usepackage{hyperref}

\newcommand{\Ham}{\hat{H}} % Hamiltonien
\renewcommand{\vec}{\mathbf} % Vecteur
\newcommand{\e}[1]{\text{e}^{#1}} % exponentielle
\newcommand{\eps}{\varepsilon} % epsilon
\graphicspath{{./images/},{./}}

\begin{document}

\title{A graphene inspired electromagnetic superlens}

\author{Sylvain Lanneb\`{e}re\textsuperscript{1}}
\author{M\'{a}rio G. Silveirinha\textsuperscript{2}}
\email{To whom correspondence should be addressed:
mario.silveirinha@co.it.pt}
 \affiliation{\textsuperscript{1}
Department of Electrical Engineering, University of Coimbra and
Instituto de Telecomunica\c{c}\~{o}es, 3030-290 Coimbra, Portugal}
\affiliation{\textsuperscript{2}University of Lisbon -- Instituto
Superior T\'ecnico and Instituto de Telecomunica\c{c}\~{o}es, 1049-001
Lisboa, Portugal}

\date{\today}

\begin{abstract}
In this paper we propose a new paradigm to create  superlenses inspired by \textit{n}-\textit{p}-\textit{n} junctions of graphene. We show that by adjoining a \textit{n}-type region and a \textit{p}-type region with a crystal dislocation, it is possible to mimic the interaction of complementary Hamiltonians and achieve subwavelength imaging. We introduce an effective model of the system, and show that it predicts perfect lensing for both propagating and evanescent waves due to the excitation of a resonant mode at the interface between each region. This phenomenon is the consequence of a nontrivial boundary condition at the \textit{n}-\textit{p} interfaces due to a dislocation of the graphene ``atoms''.
We discuss practical realizations of such superlenses in electronic and photonic platforms. Using full wave simulations, we study in detail the performance of a photonic realization of the lens based on a  honeycomb array of dielectric cylinders embedded in a metal.
\end{abstract}

\maketitle

\section{Introduction}

The ability to transport the evanescent part of the electromagnetic waves is the key ingredient to create perfect imaging devices with a resolution that is not limited by diffraction. In that respect, the groundbreaking discovery that a slab of material with simultaneously negative permittivity and permeability, the so-called superlens \cite{pendry_negative_2000}, is able to focus not only the far-field but also the near-field components of the electromagnetic field, generated a strong excitement in the scientific community \cite{pendry_negative_2004,zhang_superlenses_2008,capolino_theory_2009,solymar_waves_2009}.
Subsequently, a plethora of new ideas to mimic  or approximate the behavior of this ideal lens were developed not only for electromagnetic  waves \cite{alu_pairing_2003,fang_diffraction-limited_2005,verhagen_three-dimensional_2010,luo_all-angle_2002,luo_negative_2003,luo_subwavelength_2003,cubukcu_subwavelength_2003,shin_all-angle2_2006,fan_all-angle_2006,scalora_negative_2007,tretyakov_waves_2003,pendry_chiral_2004,wu_theory_2010,pendry_focusing_2003,maslovski_phase_2003,maslovski_perfect_2012,pendry_time_2008,freire_planar_2005,silveirinha_superlens_2008,shin_all-angle_2006} but also for electronic \cite{kobayashi_complementary_2006,silveirinha_effective_2012,silveirinha_spatial_2013,hrebikova_perfect_2014} and acoustic waves  \cite{li_double-negative_2004,ambati_surface_2007}.

Here, by combining the concepts of complementary materials \cite{pendry_focusing_2003,kobayashi_complementary_2006,silveirinha_spatial_2013} and of graphene \textit{n}-\textit{p}-\textit{n} junctions \cite{cheianov_focusing_2007}, we propose a new paradigm to realize photonic superlenses. First, we  show in section \ref{sec:electronic_superlens} that a graphene \textit{n}-\textit{p}-\textit{n} junction with a dislocation at the \textit{n}-\textit{p} interfaces may enable a perfect electron tunneling for all incident angles.
Then, in section \ref{sec:photonic}, by using a strict analogy between the two dimensional (2D) Schr\"{o}dinger and Maxwell equations, we extend this scheme to photonics and propose a specific realization based on dielectrics and metals. We study numerically the performance of the proposed lens and demonstrate that its resolution is ultimately determined by the lattice constant of the crystal.

\section{Graphene inspired electronic superlens} \label{sec:electronic_superlens}

Generally, perfect lensing relies on two effects: negative refraction, to focus the diverging rays, and amplification of evanescent waves, to restore the subwavelength spectrum \cite{pendry_negative_2000,maslovski_phase_2003}. In this section we combine the concepts of  \textit{n}-\textit{p} junction \cite{cheianov_focusing_2007} and of complementary materials \cite{pendry_focusing_2003,kobayashi_complementary_2006,silveirinha_spatial_2013}, and describe a general scheme to create ``perfect" lenses for electron waves.

\subsection{Electronic superlens based on a graphene \textit{n}-\textit{p}-\textit{n} junction}

\begin{figure}[!ht]
\centering
\includegraphics[width=.95\linewidth]{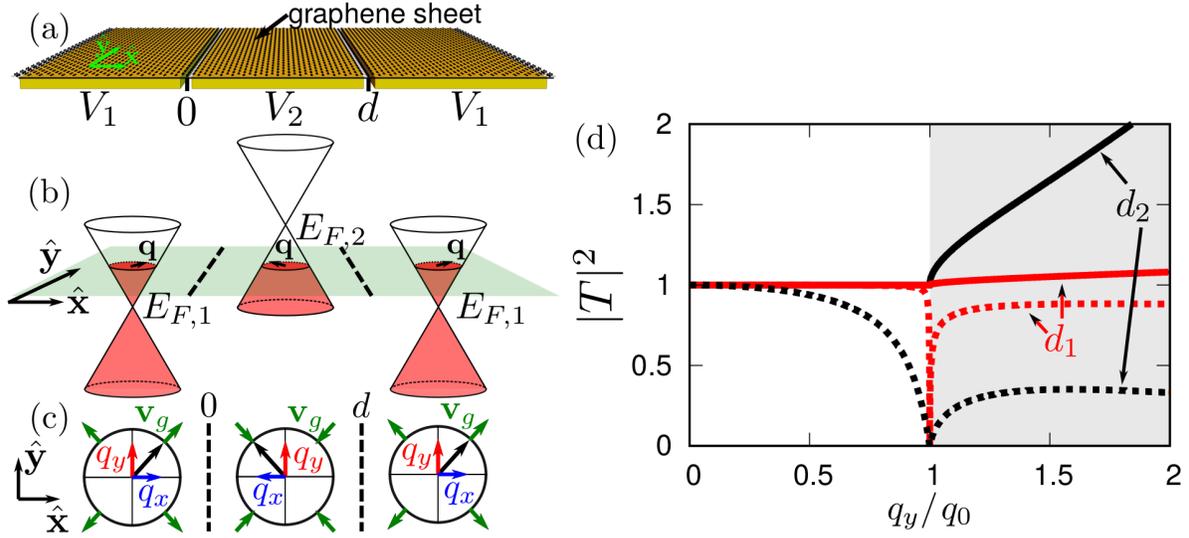}
         \caption{(a) Schematic of an electronic lens of thickness $d$ made of a graphene \textit{n}-\textit{p}-\textit{n} junction as proposed in \cite{cheianov_focusing_2007}. The graphene sheet is on top of three electrodes with potentials $V_1$ (\textit{n}-doped regions) and $V_2$ (\textit{p}-doped region). (b) Energy band diagram (Dirac cones centered at $K$ or $K'$) in each section of the lens for electrons with a fixed energy $E=E_0$. Here, $\vec{q}=q_x\hat{\vec{x}}+q_y\hat{\vec{y}}$ is the wavevector measured relatively to the $K$ or $K'$ point. (c) Contours of constant energy for the situation depicted in (b). The wavevectors in the different regions are represented by the black arrows, and their projections on the coordinate axes by the red and blue arrows. The green arrows indicate the group velocity. The conservation of the parallel wavevector $q_y$ at the interfaces and the choice of a transmitted wave with group velocity oriented towards the $+x$ direction imply a
negative refraction. (d) Transmissivity $|T|^2$ of the \textit{n}-\textit{p}-\textit{n} junction  as a function of the normalized parallel wavevector $q_y$ with $q_0\equiv\frac{|E_{F,2}-E_{F,1}|}{2\hbar v_F}$ at the energy  $E=E_0$ for two thicknesses of the \textit{p}-doped region ($d_2>d_1$). Dashed lines: Cheianov lens represented in Fig. \ref{fig:cheianov_lens} (a). Solid lines: superlens represented in Fig. \ref{fig:system_implemented} with  the \textit{n}-type and \textit{p}-type regions behaving as complementary materials. In both cases, the structure is assumed infinite along the $y$ direction. The thicknesses used are $d_1=2\lambda_0$ and $d_2=20\lambda_0$ with $\lambda_0=2\pi/|K|$ the wavelength associated with the $K$ or $K'$ points. }
         \label{fig:cheianov_lens}
\end{figure}

As shown in \cite{cheianov_focusing_2007}, an incident electron wave propagating towards the boundary between the \textit{n}- and \textit{p}-doped graphene regions \cite{novoselov_electric_2004} can be negatively refracted at the interface. As illustrated in Fig. \ref{fig:cheianov_lens}(a)--(c) for a \textit{n}-\textit{p}-\textit{n} junction, the negative refraction is a consequence of two properties: the isotropy of the energy dispersion close to the Dirac cones and the fact that the electrons propagating in the \textit{n}-doped regions belong to the conduction band whereas the electrons propagating in the \textit{p}-doped region belong to the valence band.
For an incoming wave with energy $E$ exactly equidistant from the Fermi energies $E_{F,1}$ and $E_{F,2}$ of the \textit{n}- and \textit{p}-doped regions, i.e. for $E=E_0$ with $E_0 \equiv(E_{F,1}+E_{F,2})/2$, a thin \textit{n}-\textit{p}-\textit{n} junction can focus almost perfectly an incoming plane wave for any incidence angle up to grazing incidence as shown in Fig. \ref{fig:cheianov_lens} (d) (dashed red curve), due to two successive negative refractions at the interfaces. 
Unfortunately, despite these interesting features, the imaging properties of the \textit{n}-\textit{p}-\textit{n} junction, hereafter referred to as a "Cheianov lens", are quickly degraded as soon as the thickness of the lens (i.e. of the \textit{p}-doped region) is a significant fraction of the wavelength. In fact, for large values of $d$, the transmissivity of the lens is near unity only for incidence angles close to normal incidence (see the dashed black curve in Fig. \ref{fig:cheianov_lens} (d)). Furthermore, the Cheianov lens is unable to focus evanescent waves. Thus, its resolution is limited by diffraction.

An interesting solution for this problem based on complementary graphene superlattices was proposed some time ago \cite{silveirinha_effective_2012, fernandes_wormhole_2014}. In this system, the electrons in the $j$-th superlattice behave as massless particles described by a generalized Dirac Hamiltonian
\begin{align}\label{E:Hamiltonian}
 \Ham_j =-i\hbar v_F\left(\sigma_x \hat{\vec{x}}+ \chi_j \sigma_y \hat{\vec{y}} \right)\cdot \nabla + V_j
 \end{align}
where $V_j$ is a constant potential, $\sigma_x$, $\sigma_y$ are the Pauli matrices, $v_{F}$ is the Fermi velocity of graphene (assumed to be independent of the doping level) and $\chi_j$ is the anisotropy parameter. The parameter $\chi_j$ is typically less than 1 in absolute value because in a superlattice the electron wave can propagate with different velocities along the $x$ and $y$ directions. A perfect electron tunneling can be achieved with a lens analogous to the \textit{n}-\textit{p}-\textit{n} junction described above and made of two complementary superlattices $A$ and $B$ such that $\chi_A=-\chi_B$ \cite{silveirinha_effective_2012}.\\
Remarkably, the Hamiltonian of graphene is also given by Eq. \eqref{E:Hamiltonian} with $\chi_j=1$ at the $K$ point and $\chi_j=-1$ at $K'$. However, even though the required material complementarity naturally occurs for the Dirac electrons of graphene \cite{silveirinha_spatial_2013},  the coupling between waves propagating near the $K$ and $K'$ points is negligible.  Interestingly, as described next, it is possible to effectively implement the complementary material with $\chi=-1$ in graphene, simply by tailoring the boundary conditions at the interfaces of the \textit{n}-\textit{p}-\textit{n} junction.\\
To explain this idea, we start by noting that the standard boundary condition in graphene involves the continuity of the pseudo-spinor $\psi_j=\left( \psi_{1j} \quad \psi_{2j} \right)^T$ at the two sides of an interface: ${\psi _{1A}} = {\psi _{1B}},\,\,{\psi _{2A}} = {\psi _{2B}}$.
Suppose however that somehow one can engineer a boundary condition of the form ${\psi _{1A}} = {\psi _{2B}},\,\,{\psi _{2A}} = {\psi _{1B}}$ so that the first (second) component of the pseudospinor in region A is identical to the second (first) component of the pseudo-spinor in region B. This boundary condition can be written in a compact form as:
\begin{equation}\label{E:boundary_condition}
\psi _A = \sigma _x  \psi _B.
\end{equation}
Note that this boundary condition ensures the continuity of the probability current \cite{fernandes_wormhole_2014}.
It can be easily shown that the boundary condition \eqref{E:boundary_condition} is equivalent to redefine the Hamiltonian in region B as ${{\hat H}_{B,{\rm{ef}}}} = {\sigma _x}{{\hat H}_B}{\sigma _x}$, and combined with the continuity of the pseudo-spinor at the interface. But it is simple to check that if ${{\hat H}_B}$ is the standard Hamiltonian of graphene near the $K$ point with $\chi=1$, then  ${{\hat H}_{B,{\rm{ef}}}}$ corresponds exactly to the complementary Hamiltonian with $\chi=-1$. Thereby, it follows that by tailoring the boundary condition satisfied by the pseudo-spinor as in Eq. \eqref{E:boundary_condition} it is possible to effectively emulate the interface between two complementary regions, even though the two regions are effectively identical (i.e., are the same material) apart from the applied gate voltages.

\begin{figure}[!ht]
\centering
\includegraphics[width=.6\linewidth]{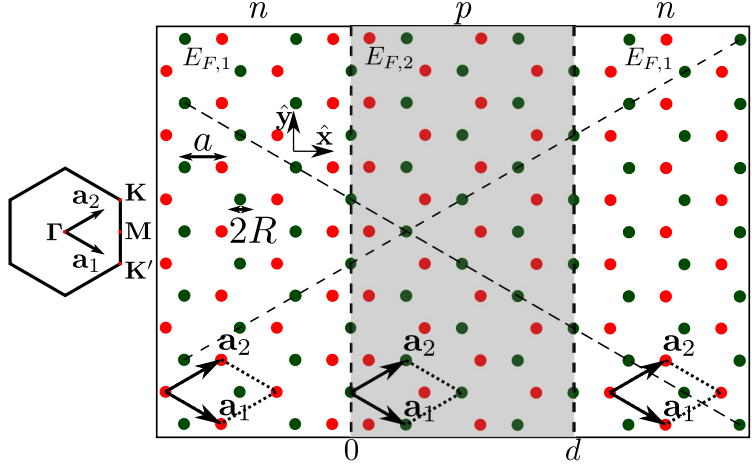}
         \caption{Schematic of the \textit{n}-\textit{p}-\textit{n} junction with crystal dislocations at the interfaces. The dislocations effectively interchange the two graphene sublattices at  interfaces (represented by the red and green circles). 
         Due to the dislocations, the \textit{p}-type region effectively behaves as the complementary material of the \textit{n}-type region. Within the effective model approximation, the system behaves as a perfect lens. The different background colors indicate the different gate voltages applied to each region of the junction.  The first Brillouin zone together with the points of high symmetry of the honeycomb lattice are represented on the left.}
         \label{fig:system_implemented}
\end{figure}

In practice, the desired boundary condition requires that the roles of sublattices 1 and 2 are interchanged in the \textit{n}- and \textit{p}-type regions. This suggests that the interface should be such that the atoms on the interface should belong to different sublattices of the \textit{n}- and \textit{p}-type regions. This can be done with the construction represented in Fig. \ref{fig:system_implemented}. As seen, the scattering centers on the interfaces are shared by the crystalline structures in regions A and B, but are associated with different sublattices. From a technical standpoint,  each of the interfaces is characterized by a dislocation \cite{ashcroft_solid_1976}.
Thus, with suitable crystal dislocations at the interfaces, it may be possible to transform a regular \textit{n}-\textit{p}-\textit{n} junction into a superlens.
\\
It is curious to point out that the wave propagation along $K$ or $K'$ in a honeycomb lattice can be discriminated by looking at the relative position of the two constituents of the unit cell. By following the two oblique dashed lines passing through the centers of the green disks in  Fig. \ref{fig:system_implemented}, one can see that the nearest neighbors are located below (above) the dashed line in the \textit{n}-doped regions and above (below) it in the \textit{p}-doped regions, as if the meaning of $K$ and $K'$ is interchanged when an interface is crossed. This gives a different perspective on the reason why the \textit{p}-type region effectively emulates (due to the dislocations) the Hamiltonian of the $K'$ point.
%
%meaning that effectively %a wave initially %propagating close to $K$ %($K'$) in the %\textit{n}-doped region, %propagates close to $K'$ %($K$) in the %\textit{p}-doped %region.
Finally, it should be mentioned that the construction of Fig. \ref{fig:system_implemented} is not unique. In principle, other dislocations that effectively interchange the sublattices of the  \textit{n}- and \textit{p}-doped regions at the interfaces,
may also be suitable to implement a superlens.\\

The transmissivity  of the \textit{n}-\textit{p}-\textit{n} junction with the dislocations can be found by writing the electronic wave function $\psi$ in each region and imposing the boundary condition \eqref{E:boundary_condition}. Assuming without loss of generality that the propagation is close to $K$ point, it follows that the wavefunction in each region is \cite{katsnelson_physics_2020}
\begin{align} \label{E:psi_effective_medium}
\psi(x,y)=\dfrac{\e{i   q_y  y  }}{\sqrt{2}}
\begin{cases}
   \begin{pmatrix} 1  \\  \e{i\theta_{q_1}}   \end{pmatrix} \e{i   q_{x,1}  x  } + R  \begin{pmatrix} 1  \\  -\e{-i\theta_{q_1}}     \end{pmatrix} \e{-i q_{x,1} x }, & \quad x\leq0 \\
 A_1 \begin{pmatrix} 1  \\  -\e{i\theta_{q_2}}\end{pmatrix} \e{i   q_{x,2}  x  } +   A_2   \begin{pmatrix} 1  \\  \e{-i\theta_{q_2}}    \end{pmatrix} \e{-i q_{x,2}  x  }  , & \quad 0\leq x \leq d\\
 T      \begin{pmatrix} 1  \\  \e{i\theta_{q_1}}    \end{pmatrix} \e{i  q_{x,1}   (x-d) } , & \quad x\geq d
\end{cases}
\end{align}
%
% %
% \begin{align}
% \psi(x,y)=\dfrac{\e{i   q_y  y  }}{\sqrt{2}}
% \begin{cases}
%    \begin{pmatrix} 1  \\  \frac{q_1}{q_{x,1}-iq_y}   \end{pmatrix} \e{i   q_{x,1}  x  } + R  \begin{pmatrix} 1  \\  \frac{q_1}{-q_{x,1}-iq_y}   \end{pmatrix} \e{-i q_{x,1} x }, & \quad x\leq0 \\
%  A_1 \begin{pmatrix} 1  \\  \frac{-q_2}{q_{x,2}+iq_y}   \end{pmatrix} \e{i   q_{x,2}  x  } +   A_2   \begin{pmatrix} 1  \\  \frac{-q_2}{-q_{x,2}+iq_y}   \end{pmatrix} \e{-i q_{x,2}  x  }  , & \quad 0\leq x \leq d\\
%  T      \begin{pmatrix} 1  \\  \frac{q_1}{q_{x,1}-iq_y}   \end{pmatrix} \e{i  q_{x,1}   (x-d) } , & \quad x\geq d
% \end{cases}
% \end{align}
% %
%
%$\tan\left( \theta_{q_i}\right)=q_y/q_{x,i}$
where $\theta_{q_i}=\arg \left( {{q_{x,i}} + i{q_y}} \right)$
, $q_{x,i}^2=q_i^2-q_y^2$ with $q_i=\frac{\left|E-E_{F,i}\right|}{ \hbar v_F }$, $R$ and $T$ are the reflection and transmission coefficients, and $A_1$ and $A_2$ are the amplitude of the waves in the \textit{p}-doped region. By enforcing the
boundary condition (\ref{E:boundary_condition}) at the interfaces, it is found that the transmissivity of the lens is:
\begin{align}\label{E:transmission}
\left| T \right|^2=\left|\frac{1}{\cos (q_{x,2} d)+i \sin (q_{x,2} d)  \frac{q_1 q_2- q_y^2 }{q_{x,1} q_{x,2}}} \right|^2
\end{align}
This is exactly the same result that would be obtained with the continuity of pseudospinor and with a \textit{p}-type region described by the Dirac Hamiltonian of the $K'$ point.

Remarkably for electron states with energy $E=E_0$, one has $q_1=q_2$ and the transmissivity simplifies exactly to $\left| T \right|^2=\left| \e{-i q_{x,1} d} \right|^2$ \cite{silveirinha_effective_2012}. For this energy the propagating electron waves are perfectly transmitted by the lens whereas the evanescent waves are exponentially amplified with a growing rate proportional to the thickness of the lens. These properties are illustrated in Fig. \ref{fig:cheianov_lens} (d) (solid lines) for two different values of the lens thickness $d$. Thus, regardless of the thickness, the amplification compensates the evanescent decay of the waves outside the \textit{p}-doped region and allows for a perfect imaging.

It should be noted that a related electronic superlens made of complementary graphene ribbons was studied in Ref. \cite{kobayashi_complementary_2006} by Kobayashi based on a different theoretical framework. In particular, Kobayashi formalism leads to lenses similar to the one shown in Fig. \ref{fig:system_implemented}. However, different from our solution, the  Kobayashi superlens is not coupled to a \textit{n}-\textit{p}-\textit{n} junction and consequently only works for electron states with energy coincident with the Dirac point, which is of limited interest because the density of electronic states vanishes at the Dirac point; furthermore, the Dirac point is a singular point as it may give rise to conical diffraction \cite{peleg_conical_2007}. These problems are solved by our solution which is a modification of the \textit{n}-\textit{p}-\textit{n} junction.

%\subsection{Practical implementation in artificial graphene}
In graphene each carbon atom is linked to three neighbors by three covalent bonds. Thereby, interfaces such as the ones materialized by the dashed vertical lines in Fig. \ref{fig:system_implemented} or those considered in Ref. \cite{kobayashi_complementary_2006} may be chemically unstable. Fortunately, the scheme described above is applicable not only to graphene but also to any system described by a Dirac Hamiltonian. In particular, any platform that mimics the properties of graphene \cite{polini_artificial_2013} --hereafter referred to as artificial graphene-- and that enables creating interfaces as in Fig. \ref{fig:system_implemented} is a suitable candidate to realize the superlens.
An electronic implementation of the lens could, for example, rely on a 2D electron gas (2DEG) modulated by a periodic electrostatic potential $V(\vec{r})$ with the honeycomb symmetry \cite{gibertini_engineering_2009,lannebere_effective_2015,wang_observation_2018}.  Furthermore, the proposed superlens is not restricted to electrons but can as well be implemented with electromagnetic or matter waves \cite{polini_artificial_2013}.
Motivated by this possibility, in the next section we investigate a photonic realization of the superlens.

\section{Photonic implementation}  \label{sec:photonic}
In a previous work, we have shown that the wave propagation in artificial graphene can be exactly mimicked by a particular photonic crystal formed by air rods in a metallic background \cite{lannebere_photonic_2019}. Inspired by this result, and by its relative simplicity and practical interest, next we propose the realization of a photonic graphene-type superlens.\\
To do so, we use a formal analogy between the 2D Schr\"odinger and Maxwell equations to transpose to photonics the ideas developed thus far.

\subsection{Photonic artificial graphene} \label{sec:photonic_artificial_graphene}
The photonic graphene introduced in Ref. \cite{lannebere_photonic_2019} consists of honeycomb array of air cylinders embedded in a metallic background. Our solution explores an analogy between the 2D Schr\"odinger and Maxwell equations, which for convenience is reviewed next.

The analogy uses as a starting point  a particular realization of artificial graphene in a 2D electron gas (2DEG) patterned with scattering centers \cite{gibertini_engineering_2009,lannebere_effective_2015,wang_observation_2018}. The electrons in this system are microscopically described by a Hamiltonian $\Ham_\text{mic}= \frac{\hat{\vec{p}}^2}{2m_b}    + V(\vec{r})$ where $V(\vec{r})$ is a periodic potential with the honeycomb symmetry, $m_b$ is the electron effective mass and $\hat{\vec{p}}=-i\hbar\nabla$ the momentum operator. The stationary states $\psi$ with energy $E$ are solutions of the time-independent Schr\"odinger's equation $\Ham_\text{mic}\psi=E \psi$ in the 2DEG:
\begin{align} \label{E:Schrodinger_artificial_graphene}
 \left[ \frac{\hbar^2}{2m_b}  \nabla^2 - V(\vec{r}) + E \right] \psi =0.
 \end{align}
With a suitable $V$ with the honeycomb symmetry this system behaves effectively as graphene \cite{gibertini_engineering_2009,lannebere_effective_2015,wang_observation_2018}.\\
To obtain an electromagnetic equivalent of this platform, we consider a 2D ($\partial/\partial z=0$) nonmagnetic photonic crystal with a relative permittivity described by a Drude model of the form
\begin{align}   \label{E:Drude_permittivity}
\eps(\vec{r})&=1-\frac{\omega_p^2(\vec{r}) }{\omega^2},
\end{align}
where $\omega_p(\vec{r})$ is the spatially dependent plasma frequency of the metal.
For stationary states, in the absence of sources, the wave equation for TE-waves ($\vec{E}=E_z \hat{\vec{z}}$) in this system reduces to $\left[  \nabla^2 + \frac{\omega^2}{c^2} \eps(\vec{r})\right]  E_z \hat{\vec{z}} = 0$. Thus, upon substitution of \eqref{E:Drude_permittivity}, one finds that:
\begin{align}  \label{E:Maxwell_artificial_graphene_TE}
\left[ \nabla^2 + \frac{\omega^2}{c^2}    -  \frac{\omega_p^2(\vec{r})}{c^2}\right]  E_z \hat{\vec{z}}   &=     0.
\end{align}
A direct comparison between equations \eqref{E:Schrodinger_artificial_graphene} and \eqref{E:Maxwell_artificial_graphene_TE} shows that their solutions can be precisely matched by taking
\begin{subequations}\label{E:equival_electro_photonic}
\begin{align}
\frac{2 m_b E }{\hbar^2} &= \frac{\omega^2}{c^2},  \label{E:equival_E_omega}\\
\frac{2m_b V(\vec{r})}{\hbar^2}   &= \frac{\omega_p^2(\vec{r})}{c^2} \label{E:equival_V_omegaP}.
\end{align}
\end{subequations}
Thus, it follows that for TE-waves a metal with a periodically modulated plasma frequency is a photonic equivalent of graphene. The periodic modulation has the honeycomb symmetry. As discussed in \cite{lannebere_photonic_2019}, this analogy can be used for any modulated 2DEG provided  the electrostatic potential $V(\vec{r})$ only assumes positive values. Moreover, for negative potentials bounded from below, it is always possible to shift the energy origin in such a manner that the potential function becomes positive, and thus, in such a case, the analogy can also be used.\\
Interestingly, our system can be implemented in practice by enclosing the photonic crystal with two metallic (ideally perfectly conducting) plates separated by some distance $h$ (the air rods are  perpendicular to the plates). Provided $h$ is a small fraction of the wavelength the only mode that propagates in such a parallel-plate waveguide is the TE mode considered in our analysis. Thus, for a small separation of the walls with respect to the wavelength, a truncated photonic crystal may behave as a (quasi-)2D photonic implementation of graphene. Note that in graphene the electron orbitals are not strictly in a plane as there is some spreading of the wave function along the $z$-direction, somewhat analogous to the parallel-plate guide electromagnetic realization.

\subsection{Photonic graphene superlens}

Now that a photonic equivalent of graphene was identified, the next step to realize the superlens is to find an electromagnetic equivalent of the \textit{n}-\textit{p}-\textit{n} junction of Fig. \ref{fig:system_implemented}. Creating an \textit{n}-\textit{p}-\textit{n} junction, involves nothing more than shifting the value of the microscopic $V$ in the entire \textit{p}-region by a certain offset determined by the gate voltage in the \textit{p}-region. In the photonic design, this leads to a global shift of the plasma frequency in the \textit{p}-region. 
\\

To illustrate these ideas, we recover a particular artificial graphene design considered in previous works \cite{gibertini_engineering_2009,lannebere_effective_2015}. It consists of a modulated 2DEG with the honeycomb symmetry and a lattice constant $a=150 \,\mathrm{nm}$ and $R=52.5\, \mathrm{nm}$ (see Fig. \ref{fig:system_implemented}). The electrostatic potentials in the \textit{n}-doped regions are taken as $V=0.8 \,\mathrm{meV}$ inside the disks of radius $R$ and $V=0$ outside. In order to implement the junctions,  $V$ suffers an additional offset (both inside and outside the scattering centers) of $+6 \,\mu\mathrm{eV}$ in the \textit{p}-doped region as compared to the \textit{n}-doped regions. All the numerical examples of the paper are based on this specific geometry but it is important to note that all the design parameters can be rescaled by choosing a different lattice constant $a$.\\

The photonic design is obtained using the transformation \eqref{E:equival_electro_photonic}. Such a construction implies that the photonic analogue of the \textit{n}-doped artificial graphene is a photonic crystal formed by an honeycomb array of air cylinders embedded in a metallic background with plasma frequency $\omega_p a/c \approx 5.626$ ($\omega_p$ is in the UV). Similarly, the photonic analogue of the \textit{p}-doped artificial graphene is a photonic crystal with the same geometry, formed by metallic cylinders with plasma frequency $\omega_p a/c\approx 0.48724$ (infrared range) embedded in another metal with plasma frequency $\omega_p a/c\approx 5.647$ (UV). \\

\begin{figure}[!ht]
\centering
\includegraphics[width=.55\linewidth]{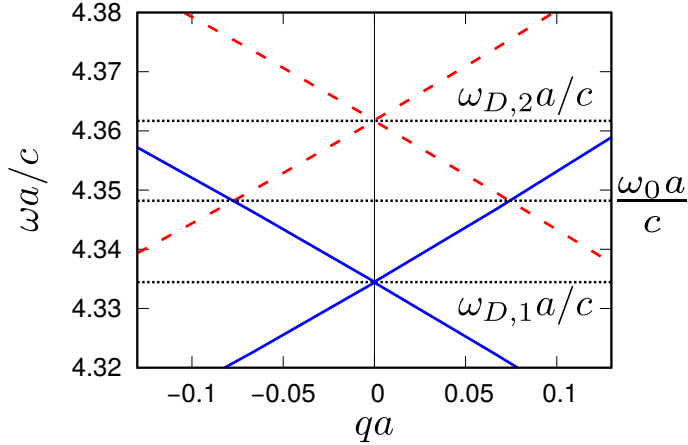}
         \caption{Photonic band diagrams near the $K$ (or $K'$) point in the photonic analogue of the \textit{n}-doped (solid blue lines) and \textit{p}-doped (dashed red lines) regions. The normalized frequencies for the Dirac cones origin $\omega_{D,i}$ ($i=1,2$) are marked by horizontal dotted lines. The effective medium model predicts perfect imaging at the frequency $\omega_0$ (dotted line in the center). }
         \label{fig:BD_photonic_npnJunction}
\end{figure}

Figure \ref{fig:BD_photonic_npnJunction}  shows the frequency dispersion of the electromagnetic modes in each photonic crystal, computed with the plane wave method \cite{joannopoulos_photonic_2008} using 121 harmonics.
As seen, the photonic dispersion near the $K$ and $K'$ points consists of a Dirac cone confirming that each region of the photonic lens indeed mimics graphene. Moreover, the Dirac cones have a similar shape (similar group velocities) and their origins, located at the Dirac frequencies $\omega_{D,i}$ with $i=1,2$, are slightly shifted in frequency. Consequently, as seen in section \ref{sec:electronic_superlens}, a \textit{n}-\textit{p}-\textit{n} junction of the two crystals, with suitable crystal dislocations at the interfaces, should provide negative refraction and subwavelength imaging at the frequency $\omega_0\equiv(\omega_{D,1}+\omega_{D,2})/2$. The crystal dislocations ensure that the roles of the sublattices in the different regions are interchanged.

\subsection{Numerical results} \label{sec:numerical_results}
\begin{figure}[!ht]
\centering
\includegraphics[width=.6\linewidth]{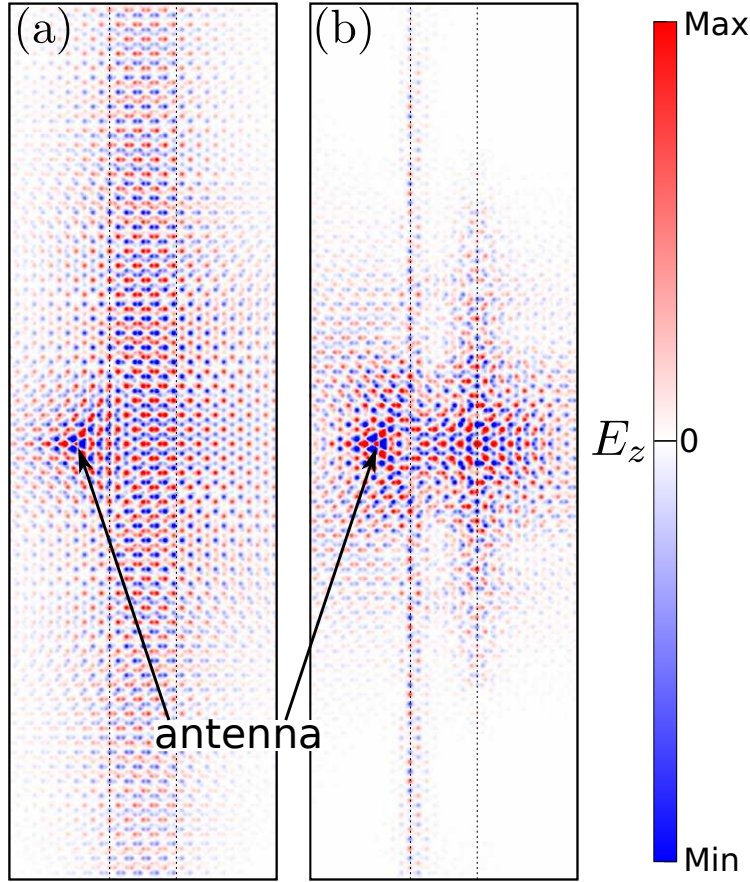}
         \caption{Time snapshot of the electric field $E_z$ emitted by a short dipole antenna at $\omega \approx 0.998 ~ \omega_0$ in  (a) a Cheianov  \textit{n}-\textit{p}-\textit{n} lens (no crystal dislocations)  (b) \textit{n}-\textit{p}-\textit{n} superlens with crystal dislocations. The vertical dashed lines marks the boundaries of the \textit{p}-doped region.}
         \label{fig:CST_compa_n_p_n_junction}
\end{figure}

\begin{figure}[!ht]
\centering
\includegraphics[width=0.95\linewidth]{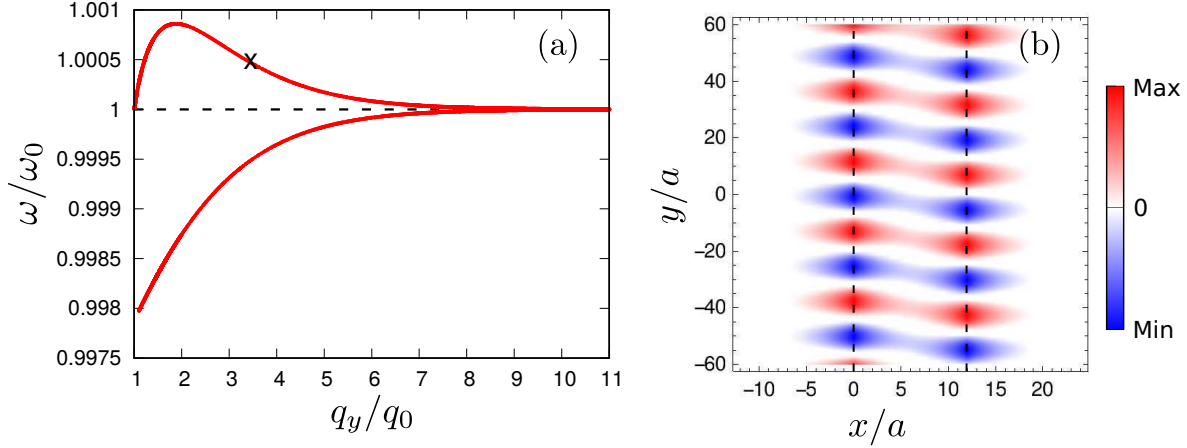}
         \caption{(a) Frequency dispersion of the surface states of the electromagnetic superlens of Fig. \ref{fig:field_on_line}(b) calculated with the effective medium theory. (b) Plot of $\mathrm{Re}\left\{\psi_1(x,y)+\psi_2(x,y) \right\}$  for the point marked by a black cross in panel (a)  [corresponding to  $q_y=3.345q_0$ and $\omega=1.0005\omega_0 $  ]. The vertical dashed lines mark the boundaries of the \textit{p}-doped region at $x=0$  and $x=12a$. }
         \label{fig:fig5}
\end{figure}

To verify these predictions, we used CST Microwave Studio to make a full-wave time-domain simulation of the photonic \textit{n}-\textit{p}-\textit{n} junction response.
We considered two cases: a \textit{n}-\textit{p}-\textit{n} junction with no dislocations (Fig. \ref{fig:CST_compa_n_p_n_junction} (a)), and a \textit{n}-\textit{p}-\textit{n} junction with dislocations that interchange the crystal sublattices (Fig. \ref{fig:CST_compa_n_p_n_junction} (b)). The photonic crystals are truncated at planes $z=\textit{const.}$ and terminated with perfectly conducting electric walls, analogous to the parallel plate waveguide alluded to in the end of section \ref{sec:photonic_artificial_graphene}. This guarantees that only TE waves can be excited in the system.
The structure is fed by a dipole antenna located in the \textit{n}-doped region (left-hand side) close to the interface with the \textit{p}-doped region. The antenna emits a light pulse with spectrum centered at $\omega_0$, that is taken as $\omega_0 a/c \approx  4.347 $. At frequency $\omega_0$ the values of the permittivities in the \textit{n}- and \textit{p}-doped regions  are $\eps \approx -0.675$ and $\eps \approx -0.687$ in the background regions, and $\eps =1$ and $\eps \approx 0.987$ in the cylinders, respectively. Note that the scattering centers (cylinders) behave as dielectrics at the operating frequency.\\
The electric field profiles near $\omega_0$  for the two configurations are represented in Fig. \ref{fig:CST_compa_n_p_n_junction}.
Strikingly, when the dislocations interchange the crystal sublattices the emitter can excite resonant states at the interfaces between the \textit{p}- and \textit{n}-doped regions. Using the effective model, it can be checked that a single dislocated \textit{p}-\textit{n} junction does not support any surface state at the interface. Instead, the resonant modes observed in the plot of Fig. \ref{fig:CST_compa_n_p_n_junction}(b) can be attributed to an overall resonance of the lens \cite{luo_subwavelength_2003} that occurs at the poles of the transmission coefficient given by Eq.\eqref{E:transmission}. These modes are the counterpart of the resonant modes that exist in conventional electromagnetic superlenses \cite{pendry_negative_2000,luo_subwavelength_2003}, and as such they are responsible for the amplification of evanescent waves and subwavelength imaging.  The frequency dispersion of the surface states of the photonic lens predicted by effective medium theory is represented in Fig. \ref{fig:fig5}(a).
Similar to other electromagnetic and electronic superlenses \cite{haldane_electromagnetic_2002,smith_limitations_2003,silveirinha_spatial_2013}, it is seen that in the perfectly tuned case $\omega=\omega_0$ the lens does not support interface waves ($q_y=\infty$). However, any small perturbation of the ideal case (in frequency or in the structural parameters) leads to a high quality factor interface state with finite $q_y$. From the numerical simulations we estimate that the bandwidth of the interface state is smaller than 0.3\%. It is instructive to examine the field distribution of a surface state. As a rough approximation, the microscopic wave function can be identified with the sum of the pseudo-spinor components: $\psi=\psi_1(x,y)+\psi_2(x,y)  $   [calculated using Eq.\eqref{E:psi_effective_medium}], which in the electromagnetic problem corresponds to the electric field: $\psi \to E_z$ . As shown in Fig. \ref{fig:fig5}(b), near the resonance the field is mainly concentrated close to the interfaces with a distribution similar to the field of two coupled surface plasmons (SPPs). Note that the emergence of plasmon-type modes is not trivial in our setup, because in metals plasmons occur for transverse magnetic (TM) waves, rather than for TE waves. Importantly, the effective model does not predict any surface state for a \textit{n}-\textit{p}-\textit{n} junction without dislocations, consistent with the plot of Fig. \ref{fig:CST_compa_n_p_n_junction}(a) where there is no signature of a resonant mode.
% It is clearly seen that the crystal dislocations at the interfaces have a strong repercussions on the scattering properties of the lens. Strikingly, when the dislocations interchange the crystal sublattices the emitter can excite resonant surface waves at the interfaces between the  \textit{p}- and \textit{n}-doped regions. These modes are the counterpart of the Surface Plasmon Polaritons (SPPs) in more conventional electromagnetic superlenses \cite{pendry_negative_2000,luo_subwavelength_2003}, and as such they are responsible for the amplification of evanescent waves and  subwavelength imaging. Note that the emergence of plasmon type modes is not trivial in our setup, because in metals plasmons occur for transverse magnetic (TM) waves, rather than for TE waves. Remarkably, there is no signature of surface modes for a \textit{n}-\textit{p}-\textit{n} junction without dislocations.

\begin{figure}[!ht]
\centering
\includegraphics[width=.85\linewidth]{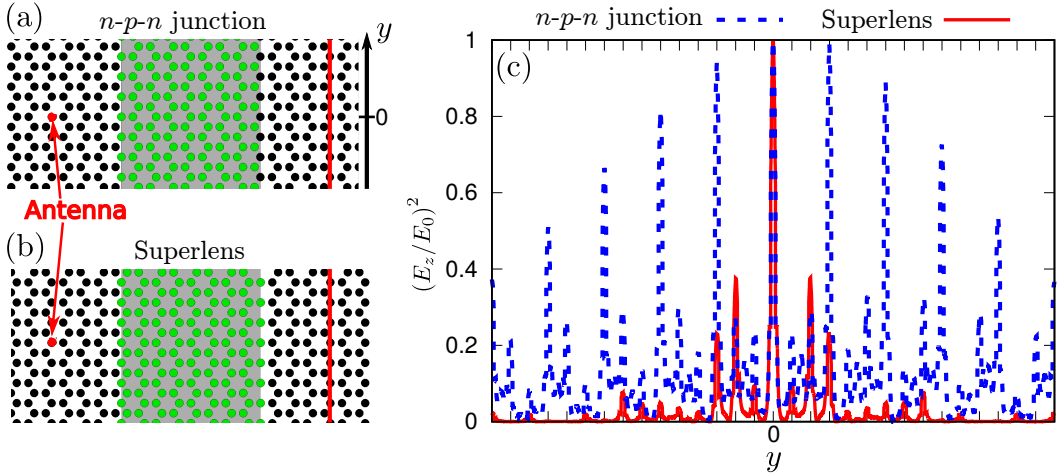}
         \caption{Top view of the central part of   (a) a Cheianov  \textit{n}-\textit{p}-\textit{n} lens (no crystal dislocations)  (b) \textit{n}-\textit{p}-\textit{n} superlens with crystal dislocations.
         The position of the antenna on the left side is marked by a red dot. (c) Normalized electric field as a function of the position $y$ on the vertical red line drawn in (a) and (b). The positions of the center of the cylinders located on this line are marked by ticks on the horizontal axis. The distance between ticks is given by the next-nearest neighbor distance ($\sqrt{3}a$).}
         \label{fig:field_on_line}
\end{figure}

Figure \ref{fig:field_on_line} shows the (full-wave) electric field profile at the image plane. The distance between the image plane and the output interface is identical to the distance between the emitter and the input interface.
As seen, the ``Cheianov lens'' (with no crystal dislocations) does not produce a clear focus whereas the superlens with crystal dislocations produces a very sharp focus. As expected, the resolution of the lens is limited by the next nearest neighbors distance. The numerical results confirm that a simple permutation of the roles of each sublattice strongly affects the scattering properties of the lens.
Of course the proposed lens suffers the same limitations as other superlenses made of photonic crystals \cite{luo_subwavelength_2003,chen_finite-size_2004,liu_limitations_2011}. Similar to other implementations of superlenses, we find that the bandwidth of the lens is rather narrow (less than 0.7\%), and thereby the lens response may be quite sensitive to small perturbations of the geometry.
 Moreover, parameters such as the lattice constant (which determines a cut-off for the evanescent spectrum) or the finite size of the lens limit the resolution. Finally, losses in the metal background also deteriorate the resolution of the superlens. Fully dielectric implementations of photonic artificial graphene \cite{ochiai_photonic_2009} may be a viable alternative to mitigate the impact of dissipation.

\section{Conclusions}
In this paper we studied graphene-type superlenses (electronic or photonic) based on  \textit{n}-\textit{p}-\textit{n} junctions. It was shown that with suitable crystal dislocations the \textit{p}-type region may effectively mimic the ideal complementary material of the \textit{n}-type region. We developed an effective medium model that predicts, within its validity, a perfect imaging. Because of the restrictions imposed by covalent bonds in graphene, we introduced an optical design of the superlens based on metallic photonic crystals. It was verified with full-wave simulations that the crystal dislocations at the interfaces effectively swap the roles of the two graphene sublattices in the  \textit{n}- and \textit{p}-type regions.
In particular, with the crystal dislocations the interfaces support surface plasmon-type excitations for TE-polarization that boost the resolution of the system, confirming the validity of the effective medium theory. The proposed scheme determines a different paradigm to create photonic superlenses based on graphene-type materials.

\section{Acknowledgement}
This work was partially funded by the Institution of Engineering and Technology (IET) under the A F Harvey
Research Prize 2018 and by
Instituto de Telecomunica\c{c}\~{o}es under Project N$^\circ$.
UID/EEA/50008/2020. S.L. acknowledges the Funda\c{c}\~{a}o para a Ci\^{e}ncia e a
Tecnologia and IT-Coimbra for the research financial support with reference DL 57/2016/CP1353/CT000.

\bibliographystyle{naturemag}

% \bibliography{Biblio}
\bibliography{Biblio_CLEAN}

\end{document}